\pdfoutput=1
\documentclass[11pt]{article}

\usepackage[preprint]{acl}

\usepackage{times}
\usepackage{latexsym}
\usepackage[T1]{fontenc}
\usepackage[utf8]{inputenc}
\usepackage{microtype}
\usepackage{inconsolata}
\usepackage{graphicx}

\usepackage{amsmath,amssymb}
\usepackage{booktabs}
\usepackage{makecell}
\usepackage{xcolor}
\usepackage{colortbl}
\definecolor{posterblue}{HTML}{75AADB}
\definecolor{posterpeach}{HTML}{FFCC99}
\definecolor{posterteal}{HTML}{99CCCC}
\definecolor{postermauve}{HTML}{CC99CC}
\definecolor{speechcol}{HTML}{3D63C9}
\definecolor{nonspeechcol}{HTML}{D9798A}
\definecolor{fitink}{HTML}{B0702A}
\definecolor{infink}{HTML}{3F6E9E}
\definecolor{clustervadshade}{HTML}{E8F3F3}
\definecolor{clustervadbwshade}{HTML}{F5E9F5}

\newcommand{\speechword}[1]{\textcolor{speechcol}{\textbf{#1}}}
\newcommand{\nonspeechword}[1]{\textcolor{nonspeechcol}{\textbf{#1}}}
\newcommand{\fitstep}[1]{\textcolor{fitink}{\textcircled{\scriptsize #1}}}
\newcommand{\stage}[1]{\textcolor{infink}{\textcircled{\scriptsize #1}}}
\usepackage{tikz}
\usetikzlibrary{arrows.meta,positioning,fit,calc,backgrounds}

\newcommand{\method}{ClusterVAD}
\newcommand{\x}{\mathbf{x}}
\newcommand{\tx}{\tilde{\mathbf{x}}}
\newcommand{\cc}{\mathbf{c}}
\newcommand{\pspeech}{P(\mathrm{sp}\,|\,\x_t)}

\title{\method{}: Threshold-Free Unsupervised Voice Activity Detection\\with Pre-Trained Neural Representations}

\author{
 \textbf{Karl El Hajal\textsuperscript{1,2}},
 \textbf{Mathew Magimai.-Doss\textsuperscript{1}}
\\
\\
 \textsuperscript{1}Idiap Research Institute, Switzerland \\
 \textsuperscript{2}EPFL, Switzerland
 \\
\\
 \texttt{
    \{karl.elhajal,mathew\}@idiap.ch
 }
}

\begin{document}
\maketitle

\begin{abstract}
Voice activity detection (VAD) is one of the first steps in speech processing pipelines, from transcription and diarization to speech language models.
Supervised detectors need labelled data and inherit the domain and annotation conventions of their training corpora. Existing unsupervised detectors rely on hand-crafted acoustic features which cannot benefit from data and degrade in far-field, noisy or atypical speech.
We propose \method{}, an unsupervised VAD built on frozen speech foundation model representations.
Fitted once on a small set of untranscribed speech, it clusters the representations and labels each cluster as speech or non-speech with sonority, an acoustic measure.
At inference, a frame-level speech probability is computed from the cluster assignments, and a two-state hidden Markov model fitted to each recording enables threshold-free detection.
Across benchmarks covering meetings, far-field and noisy recordings, Mandarin and dysarthric speech, \method{} obtains the lowest detection error on pause-level references and is competitive on speaker turn-level references.
These results indicate that speech foundation models inherently capture the distinction between speech and non-speech, which can be leveraged without training a detector.
\end{abstract}

\section{Introduction}
\label{sec:intro}

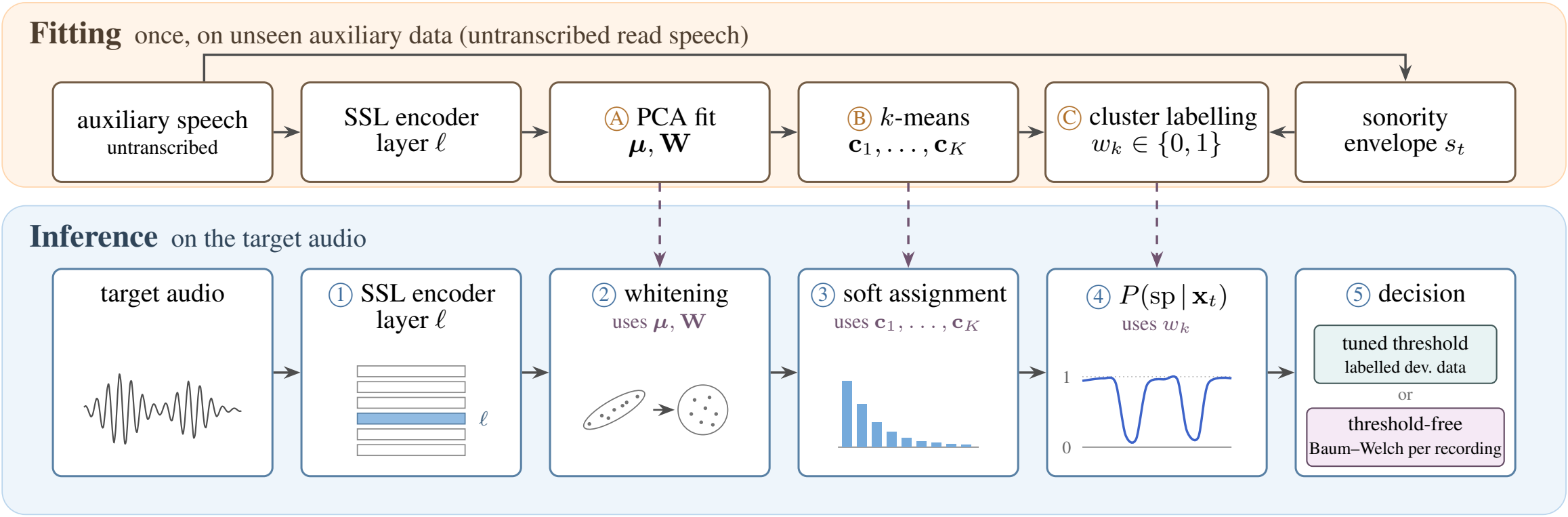
\begin{figure*}[t]
\centering
\resizebox{0.96\textwidth}{!}{\begin{tikzpicture}[
    font=\small,
    stage/.style={draw=posterblue!75!black, rounded corners=3pt, fill=white, line width=0.7pt, minimum width=26.5mm, minimum height=25mm},
    fitbox/.style={draw=posterpeach!45!black, rounded corners=3pt, fill=white, line width=0.7pt, minimum width=26.5mm, minimum height=12mm, align=center},
    ttl/.style={align=center, font=\small, inner sep=1pt},
    prm/.style={font=\scriptsize, text=postermauve!55!black},
    arr/.style={-{Stealth[length=2mm]}, line width=0.8pt, draw=black!70},
    note/.style={font=\scriptsize, text=black!60},
    uses/.style={-{Stealth[length=2mm]}, line width=0.8pt, draw=postermauve!55!black, dashed},
    choice/.style={rounded corners=2pt, line width=0.5pt, minimum width=22mm, minimum height=7mm, align=center, inner sep=0.8pt},
    ]
  \def\dx{3.0}
  \foreach \i/\n in {0/aud, 1/ssl, 2/wh, 3/sa, 4/ps, 5/dec} {
    \node[stage] (\n) at (\i*\dx, 0) {};
  }
  \node[ttl, below=1.5mm of aud.north]  {target audio};
  \node[ttl, below=1.5mm of ssl.north]  {\stage{1} SSL encoder\\[-1pt]layer $\ell$};
  \node[ttl, below=1.5mm of wh.north]   {\stage{2} whitening\\[-1pt]\textcolor{postermauve!55!black}{\scriptsize uses $\boldsymbol\mu, \mathbf{W}$}};
  \node[ttl, below=1.5mm of sa.north]   {\stage{3} soft assignment\\[-1pt]\textcolor{postermauve!55!black}{\scriptsize uses $\cc_1,\dots,\cc_K$}};
  \node[ttl, below=1.5mm of ps.north]   {\stage{4} $\pspeech$\\[-1pt]\textcolor{postermauve!55!black}{\scriptsize uses $w_k$}};
  \node[ttl, below=1.5mm of dec.north]  {\stage{5} decision};
  \node[choice, draw=posterteal!45!black, fill=clustervadshade] (thr) at ($(dec.center)+(0,0.22)$) {{\scriptsize tuned threshold}\\[-2pt]{\tiny labelled dev.\ data}};
  \node[choice, draw=postermauve!50!black, fill=clustervadbwshade] (bw) at ($(dec.center)+(0,-0.78)$) {{\scriptsize threshold-free}\\[-2pt]{\tiny Baum--Welch per recording}};
  \node[font=\tiny, text=black!55] at ($(thr.south)!0.5!(bw.north)$) {or};
  \foreach \a/\b in {aud/ssl, ssl/wh, wh/sa, sa/ps, ps/dec} { \draw[arr] (\a) -- (\b); }

  \begin{scope}[shift={($(aud.center)+(0,-0.45)$)}]
    \draw[black!70, line width=0.5pt] plot[smooth, samples=90, domain=-0.95:0.95]
      (\x, {0.42*sin(2600*\x)*(exp(-14*(\x+0.5)^2)+0.8*exp(-18*(\x-0.35)^2)+0.04)});
  \end{scope}
  \begin{scope}[shift={($(ssl.center)+(0,-0.45)$)}]
    \foreach \k in {0,...,5} {
      \ifnum\k=2 \filldraw[draw=posterblue!75!black, fill=posterblue!80] (-0.65,-0.55+\k*0.19) rectangle (0.65,-0.42+\k*0.19);
      \else \draw[black!45] (-0.65,-0.55+\k*0.19) rectangle (0.65,-0.42+\k*0.19); \fi }
    \node[font=\scriptsize, text=posterblue!60!black, anchor=west] at (0.68,-0.105) {$\ell$};
  \end{scope}
  \begin{scope}[shift={($(wh.center)+(0,-0.45)$)}]
    \draw[black!55, rotate around={30:(-0.55,0)}] (-0.55,0) ellipse (0.42 and 0.12);
    \foreach \p in {(-0.85,-0.17),(-0.7,-0.07),(-0.55,0.0),(-0.4,0.09),(-0.27,0.16),(-0.62,-0.1),(-0.45,0.05)} \fill[black!55] \p circle (0.022);
    \draw[arr, line width=0.6pt] (-0.08,0) -- (0.18,0);
    \draw[black!55] (0.52,0) circle (0.3);
    \foreach \p in {(0.42,0.12),(0.6,0.15),(0.68,-0.05),(0.5,-0.15),(0.38,-0.05),(0.55,0.02),(0.64,-0.18)} \fill[black!55] \p circle (0.022);
  \end{scope}
  \begin{scope}[shift={($(sa.center)+(0,-0.45)$)}]
    \draw[black!45] (-0.85,-0.45) -- (0.85,-0.45);
    \foreach \h [count=\k from 0] in {0.80,0.52,0.30,0.18,0.11,0.07,0.05,0.04,0.03} {
      \fill[posterblue] (-0.8+\k*0.18,-0.45) rectangle (-0.68+\k*0.18,-0.45+\h); }
  \end{scope}
  \begin{scope}[shift={($(ps.center)+(0,-0.45)$)}]
    \draw[black!45] (-0.9,-0.45) -- (0.9,-0.45);
    \draw[black!30, densely dotted] (-0.9,0.4) -- (0.9,0.4);
    \draw[speechcol, line width=0.8pt] plot[smooth] coordinates
      {(-0.9,0.35) (-0.65,0.38) (-0.5,0.33) (-0.38,-0.3) (-0.25,-0.35) (-0.12,0.3) (0.1,0.37) (0.25,0.36) (0.36,-0.25) (0.5,-0.33) (0.62,0.32) (0.9,0.38)};
    \node[font=\tiny, text=black!60, anchor=east] at (-0.9,0.4) {1};
    \node[font=\tiny, text=black!60, anchor=east] at (-0.9,-0.45) {0};
  \end{scope}
  \foreach \i/\n/\t in {0/faud/{auxiliary speech\\[-1pt]{\scriptsize untranscribed}},
                        1/fssl/{SSL encoder\\[-1pt]layer $\ell$},
                        2/fpca/{\fitstep{A} PCA fit\\[-1pt]$\boldsymbol\mu, \mathbf{W}$},
                        3/fkm/{\fitstep{B} $k$-means\\[-1pt]$\cc_1,\dots,\cc_K$},
                        4/fname/{\fitstep{C} cluster labelling\\[-1pt]$w_k \in \{0,1\}$},
                        5/fson/{sonority\\[-1pt]envelope $s_t$}} {
    \node[fitbox] (\n) at (\i*\dx, 2.9) {\t};
  }
  \foreach \a/\b in {faud/fssl, fssl/fpca, fpca/fkm, fkm/fname, fson/fname} { \draw[arr] (\a) -- (\b); }
  \draw[arr] ($(faud.north)+(0.5,0)$) -- ++(0,0.32) -| (fson.north);
  \foreach \a/\b in {fpca/wh, fkm/sa, fname/ps} { \draw[uses] (\a.south) -- (\b.north); }

  \begin{pgfonlayer}{background}
    \node[fit=(faud)(fson), inner xsep=6mm, inner ysep=5mm, yshift=4.5mm, rounded corners=8pt, fill=posterpeach!22, draw=posterpeach!80] (fitpanel) {};
    \node[fit=(aud)(dec), inner xsep=6mm, inner ysep=6mm, yshift=1.5mm, rounded corners=8pt, fill=posterblue!12, draw=posterblue!50] (infpanel) {};
  \end{pgfonlayer}
  \node[anchor=north west, font=\normalsize\bfseries, text=posterpeach!30!black] at ($(fitpanel.north west)+(2mm,-1mm)$)
    {Fitting\textnormal{\small\ \ once, on unseen auxiliary data (untranscribed read speech)}};
  \node[anchor=north west, font=\normalsize\bfseries, text=posterblue!40!black] at ($(infpanel.north west)+(2mm,-1mm)$)
    {Inference\textnormal{\small\ \ on the target audio}};
\end{tikzpicture}}
\caption{\method{} overview. \textbf{Fitting} runs once on unlabelled auxiliary data. \textbf{Inference} runs on the target audio.}
\label{fig:method}
\end{figure*}

Voice activity detection (VAD) is one of the first steps in speech and audio processing. Its goal is to detect voice segments in an audio signal and to inform subsequent processing in systems such as speech language models \citep{EndPoint_Rabiner_Sambur_1974, sharma2022moviesvad, landini2022vbx, bredin2023pyannote, kahn2020librilight, he2024emilia, nguyen2023dgslm}. Consequently, poor voice activity detection can adversely affect speech technology performance. For instance, false positives can lead a speech recognition system to yield spurious text, and false negatives can lead to the loss of information in subsequent processing. VAD has been an active area of research since the advent of telephony and remains a challenging problem. Approaches to VAD can be broadly grouped into (a) unsupervised VAD and (b) supervised VAD.

Unsupervised approaches are typically knowledge-driven. The simplest compare features such as energy, zero-crossing rate, spectral shape or pitch with a threshold \citep{EndPoint_Rabiner_Sambur_1974,ramirez2004,ghosh2011}, or fit a bi-Gaussian model to the frames of a recording and take the Gaussian with the larger mean as the speech model~\citep{bimbot2004tutorial}. Such methods work well in clean conditions but can fail in noise. Statistical model-based detectors instead model the distributions of speech and noise and apply a likelihood ratio test~\citep{sohn1999}. More recent methods encode knowledge of the voice source and vocal tract through segment-level signal-to-noise estimates \citep{tan2020rvad} or zero-frequency filtering \citep{sarkar2022zff}.

Supervised detectors instead train classifiers on labelled frames, ranging from deep belief and recurrent networks \citep{zhang2013dbn,hughes2013rnn} to convolutional models \citep{jia2021marblenet,silero2024vad} and the multi-speaker segmentation networks used in diarization \citep{bredin2021segmentation,plaquet2023powerset}. Supervised models improve with more labelled data and outperform unsupervised detectors when the test data match their training data, whereas unsupervised detectors are more reliable under domain mismatch \citep{tan2020rvad}. Meanwhile, knowledge-driven unsupervised detectors cannot leverage more data as effectively.

The present paper aims to reconcile these two directions of VAD through speech foundation model (SFM) based neural representations, which are typically learned with self-supervised learning (SSL) from large amounts of untranscribed speech.
Representations from models such as wav2vec~2.0, HuBERT, WavLM and data2vec~2.0 \citep{baevski2020wav2vec,hsu2021hubert,chen2022wavlm,baevski2023data2vec2}, have consistently been shown to yield better systems than conventional knowledge-driven features across speech processing tasks. We hypothesise that one reason is that these models should inherently learn to discriminate between speech and non-speech: to effectively abstract speech information from hundreds or thousands of hours of heterogeneous speech data, SFMs must implicitly learn to detect speech during training. Besides learning representations in a data-driven manner, this could be one of the factors that differentiate SFM representations from knowledge-driven features, where we \emph{mainly} apply knowledge about \emph{speech}. In other words, the speech-nonspeech discrimination information should be embedded in the learned representations in a form that downstream tasks can easily decipher without the need for extensive supervision or supervised training.
Indeed, in recent work \citep{vanniekerk2023urhythmic}, SFM representations have been effectively grouped into sonorants, obstruents and silence for speech rhythm modelling. Subsequent work extended this segmentation approach to convert dysarthric speech into typical speech and observed that it separates non-speech more reliably than a conventional VAD on dysarthric speech~\citep{elhajal2025rnv}. 

Inspired by these works, we propose \method{}, an unsupervised VAD built on frozen SFM representations (Figure~\ref{fig:method}).
First, a fitting stage runs once on a small set of auxiliary untranscribed speech. It clusters the representations and labels each cluster as speech or non-speech using sonority, an acoustic measure.
At inference, \method{} computes a frame-level speech probability from the cluster assignments and applies a threshold-free decision to detect speech segments.
We evaluate \method{} against supervised and unsupervised detectors on benchmarks varying in domain, language, recording conditions and annotation granularity. Results show that \method{} is robust across domains and detects pauses at a fine granularity.
Code will be released upon publication.

\section{Proposed Approach}
\label{sec:method}

\method{} treats voice activity detection as unsupervised classification of feature frames extracted with an SFM, referred to as an SSL encoder (Figure~\ref{fig:method}).
It consists of fitting steps (\fitstep{A}--\fitstep{C}), which are run once on auxiliary untranscribed speech, and inference steps (\stage{1}--\stage{5}), which can then be run on any target audio.
Fitting builds a labelled codebook: \fitstep{A} the SSL frames of the auxiliary data are whitened with PCA and \fitstep{B} clustered with $k$-means, and \fitstep{C} each cluster is labelled speech or non-speech from the sonority of its frames.
At inference, \stage{1} the SSL frames of a target recording are \stage{2} whitened and \stage{3} softly assigned to the clusters; \stage{4} the assignment is marginalised over the cluster labels into a speech probability for each frame, which \stage{5} is turned into speech segments by a decision.

\subsection{Fitting}
\label{sec:fitting}
A frozen SSL encoder maps a waveform to a sequence of frames $\x_1, \dots, \x_N \in \mathbb{R}^D$, one every 20\,ms, where $N$ is the number of frames and $D$ is the feature dimension. Features are extracted from a single intermediate layer $\ell$.
The frames of the auxiliary data form the codebook data $\mathcal{D}$.

\paragraph{\fitstep{A} PCA whitening.}
SSL layers are anisotropic: a few dimensions with large magnitudes contribute disproportionately to the norm of each frame (Appendix~\ref{app:anisotropy}), which can bias distance-based comparisons between frames towards those dimensions.
\method{} therefore centres each frame, projects it onto the leading $d$ principal components of the codebook data and scales the retained components to unit variance,
\begin{equation}
\tx_t = \boldsymbol\Lambda^{-1/2}\,\mathbf{V}^{\top}(\x_t - \boldsymbol\mu),
\label{eq:whiten}
\end{equation}
where $\boldsymbol\mu$ is the mean frame of the codebook data, and $\mathbf{V}$ and $\boldsymbol\Lambda$ hold the $d$ leading eigenvectors and eigenvalues of its covariance.

\paragraph{\fitstep{B} $k$-means codebook.}
$K$-means on the whitened frames of the codebook data gives centroids $\cc_1, \dots, \cc_K$.
The codebook assigns any frame a posterior over its clusters,
\begin{equation}
p(k\,|\,\x_t) = \frac{\exp\big(\cos(\tx_t, \cc_k)/T\big)}{\sum_{j=1}^{K} \exp\big(\cos(\tx_t, \cc_j)/T\big)},
\label{eq:assign}
\end{equation}
where $\cos(\cdot,\cdot)$ denotes cosine similarity and the temperature $T$ controls how soft the assignment is.

\begin{figure*}[t]
\begin{minipage}[t]{\columnwidth}
\centering
\includegraphics[width=0.95\columnwidth]{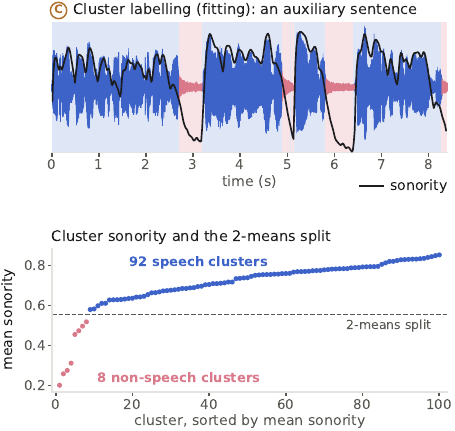}
\caption{Cluster labelling for OLIVE-J. Top: an auxiliary sentence with its sonority, each frame coloured by the label of its cluster, \speechword{speech} or \nonspeechword{non-speech}. Bottom: mean sonority of the clusters and the 2-means split.}
\label{fig:naming}
\end{minipage}\hfill
\begin{minipage}[t]{\columnwidth}
\centering
\includegraphics[width=0.95\columnwidth]{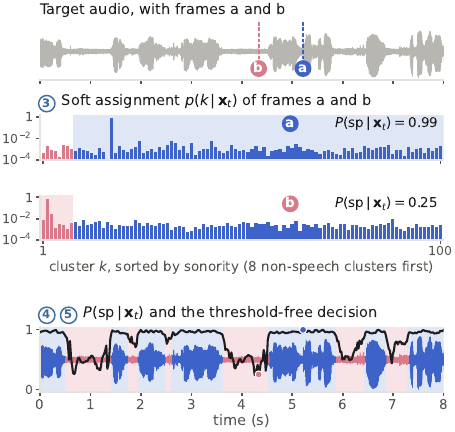}
\caption{Inference on 8\,s of an AMI meeting. \stage{3}: soft assignments of a \speechword{speech} frame and a \nonspeechword{non-speech} frame (log scale). \stage{4}--\stage{5}: speech probability curve and the threshold-free decision marked by colour shading.}
\label{fig:cascade}
\end{minipage}
\end{figure*}

\paragraph{\fitstep{C} Cluster labelling.}
The clusters need to be labelled as speech or non-speech.
\method{} labels them with the sonority envelope $s_t \in [0,1]$ of the auxiliary audio, an acoustic measure of syllabic prominence: band envelopes from a gammatone filterbank drive damped oscillators tuned to the theta rhythm of syllables (5\,Hz), and the output is summed over the most energetic bands \citep{rasanen2018}.
Sonority is typically high on vowels and voiced consonants and low in non-speech.
The average sonority of cluster $k$ is
\begin{equation}
\bar s_k = \frac{\sum_{t \in \mathcal{D}} p(k\,|\,\x_t)\, s_t}{\sum_{t \in \mathcal{D}} p(k\,|\,\x_t)}.
\label{eq:sonority}
\end{equation}
The $K$ values $\{\bar s_k\}$ are split into a low- and a high-sonority group by one-dimensional 2-means (Lloyd's algorithm with two centres), with the centres initialised at the smallest and the largest $\bar s_k$ and each cluster counted once.
The low group is labelled non-speech ($w_k = 1$) and the high group speech ($w_k = 0$).
Sonority is only computed on the auxiliary codebook data; it is not required during inference.
Figure~\ref{fig:naming} shows the split for one encoder.

\subsection{Inference}
\label{sec:inference}

Once fitting has been run on the auxiliary data, detection can be run on any target audio.

\paragraph{\stage{1}--\stage{2} Feature extraction and whitening.}
The same frozen encoder and layer $\ell$ map the input waveform to frames $\x_t$, and each frame is whitened with Eq.~\ref{eq:whiten}, using the statistics fitted in \fitstep{A}.

\paragraph{\stage{3} Soft assignment.}
\method{} computes the cosine similarity between each whitened frame and each of the $K$ centroids, and a softmax over these similarities gives the frame its posterior $p(k\,|\,\x_t)$ over the clusters (Eq.~\ref{eq:assign}).

\paragraph{\stage{4} Speech probability.}
The speech probability of a frame is the sum of the posterior probabilities assigned to the clusters labelled speech in \fitstep{C},
\begin{equation}
\pspeech = \sum_{k=1}^{K} (1 - w_k)\, p(k\,|\,\x_t),
\label{eq:pspeech}
\end{equation}
which marginalises the posterior over the cluster labels.
Long recordings are processed in overlapping windows, whose frame-level speech probabilities are averaged with Hann weights.

\paragraph{\stage{5} Decision.}
\method{} turns $\pspeech$ into speech and non-speech segments in two ways.

\emph{Tuned threshold.} A frame is labelled speech when $\pspeech$ exceeds a threshold, and the threshold and minimum segment durations (a hangover) are tuned on labelled development data.

\emph{Threshold-free decision.} \method{} can instead fit the decision to each recording, with no threshold to tune.
Let $z_t$ be the logit of $\pspeech$.
A two-state hidden Markov model has states $q_t \in \{\mathrm{ns}, \mathrm{sp}\}$, Gaussian emissions $\mathcal{N}(z_t; m_q, v_q)$ with a mean and a variance per state, and a $2 \times 2$ transition matrix.
The emissions are initialised from the distribution of $\{z_t\}$, with the non-speech Gaussian at a low percentile and the speech Gaussian at a high percentile.
Baum--Welch \citep{rabiner1989} then estimates the Gaussian parameters and the transition probabilities on the recording; the self-transitions play the role of the hangover.
Before evaluating the emissions, $z_t$ is clipped to $[m_\mathrm{ns}, m_\mathrm{sp}]$ so that the decision is monotone in the score.
Viterbi decoding finds the most probable state sequence; consecutive frames in each state form the speech and non-speech segments.
With the threshold-free decision, no decision parameter is tuned: the codebook and its cluster labels come from untranscribed auxiliary data, and the decision is fitted to each recording (Figure~\ref{fig:cascade}).
We present results with both decisions.

\section{Experimental setup}
\label{sec:setup}

\subsection{Implementation}
\label{sec:implementation}

\paragraph{SSL encoders.}
We select frozen Base and Large encoders and their layers on development data from 11 models (Appendix~\ref{app:layers}).
We use the three best performing models: HuBERT Large \citep{hsu2021hubert} at layer 10, data2vec~2.0 Base \citep{baevski2023data2vec2} at the transformer input (layer 0), and OLIVE-J \citep{elhajal2026olive} at layer 4.
OLIVE-J is a masking-only variant of OLIVE trained jointly with a waveform decoder, which also lets us synthesise audio from frames and centroids to inspect the clusters.\footnote{Audio samples: \href{https://karlhajal.github.io/ClusterVAD/}{\nolinkurl{karlhajal.github.io/ClusterVAD}}}

\paragraph{Codebook.}
The codebook data consists of 1{,}000 randomly selected utterances of LJSpeech \citep{ito2017ljspeech}, corresponding to 1.8\,h of read English from a single speaker, which none of the evaluation corpora contains. No transcripts are used.
We keep $d=128$ principal components, use $K=100$ clusters ($k$-means) and $T=0.1$ in Eq.~\ref{eq:assign}.
The sonority envelope uses 20 gammatone bands between 50 and 7{,}500\,Hz, one damped oscillator per band at 5\,Hz with $Q=0.5$, and the 10 most energetic bands \citep{rasanen2018}.
Figure~\ref{fig:naming} shows the resulting cluster labelling.
Each choice is ablated in Appendix~\ref{app:codebook}.

\paragraph{Inference.}
Recordings are processed in 5\,s windows with a 2.5\,s hop.
The speech probability is then turned into segments by one of the two decisions of Section~\ref{sec:method}.
The tuned threshold compares the speech probability with a threshold and removes speech and non-speech segments shorter than minimum durations.
The threshold-free decision is fitted on each test recording: Baum--Welch runs 10 iterations, starting from Gaussians at the 10th and 90th percentiles of $\{z_t\}$ with a quarter of its variance, equal initial state probabilities and self-transition probabilities of 0.95.

\begin{table*}[t]
\centering
\footnotesize
\setlength{\tabcolsep}{4pt}
\renewcommand{\arraystretch}{1.1}
\resizebox{\textwidth}{!}{%
\begin{tabular}{@{}llllllrr@{}}
\toprule
Dataset & Language & Domain & Noise (SNR) & Annotations & Tuning & Test files & Hours \\
\midrule
AMI headset mix & English & meetings, close-talk & -- & pause-level (forced alignment), turn-level (manual) & development set & 16 & 9.1 \\
AMI far-field & English & meetings, far-field & -- & pause-level (forced alignment), turn-level (manual) & development set & 16 & 9.1 \\
AliMeeting & Mandarin & meetings, far-field & -- & pause-level (forced alignment), turn-level (manual) & development set & 20 & 10.8 \\
Buckeye & English & interviews & -- & pause-level (hand-corrected alignment) & development set & 68 & 10.6 \\
Torgo & English & dysarthric speech & -- & pause-level (manual alignment) & leave-one-speaker-out & 662 & 0.6 \\
DIHARD III & English, Chinese & 11 domains & -- & diarization labels & two-fold & 259 & 33.0 \\
VoxConverse & English & web video & -- & diarization labels & development set & 232 & 43.5 \\
AMI + MUSAN & English & meetings, close-talk & 20 to 0\,dB & pause-level (forced alignment) & development set & 16 $\times$ 4 & 36.2 \\
LibriVAD & English & read speech & 20 to $-5$\,dB & pause-level (forced alignment) & validation set & 14{,}148 & 30.2 \\
\bottomrule
\end{tabular}}
\caption{Evaluation benchmarks. Test files: number of audio files in the test set; Hours: their total duration.}
\label{tab:datasets}
\end{table*}

\subsection{Baselines}
\label{sec:baselines}
We compare \method{} with unsupervised and supervised detectors.
The unsupervised detectors rely on signal processing: rVAD \citep{tan2020rvad}, which applies two passes of denoising and then detects speech from pitch and energy features; the pause detector of \citet{dejong2009}, which thresholds the intensity contour; WebRTC VAD \citep{webrtcvad}, which applies Gaussian mixture models to sub-band energies; and a threshold on the frame energy.
The supervised detectors are neural networks trained on labelled speech: the pyannote~3.0 segmentation model \citep{plaquet2023powerset}, which performs frame-wise powerset multi-class classification; Silero VAD \citep{silero2024vad}, a convolutional-recurrent network trained on a large multilingual corpus; MarbleNet \citep{jia2021marblenet}, a frame-level multilingual network of time-channel separable convolutions; and INA SSL-VAD \citep{pelloin-etal-2026-data}, a classifier trained on the features of a frozen data2vec~2.0 Base encoder. The latter is the supervised counterpart of \method{}: both use frozen SSL features, but INA SSL-VAD trains its classifier on labelled data.
Appendix~\ref{app:baselines} describes the training data and configuration of each baseline.

\subsection{Evaluation}
\label{sec:evaluation}

\paragraph{Datasets.}
\label{sec:datasets}
We evaluate VAD performance on benchmarks that vary in domain, language, speaker, recording conditions and noise (Table~\ref{tab:datasets}).
\textbf{AMI} \citep{carletta2005ami} and \textbf{AliMeeting} \citep{yu2022m2met} are meeting corpora in English and in Mandarin, respectively. AMI is evaluated both as a close-talk headset mix and from a far-field microphone, and AliMeeting from far-field recordings. Both come with two reference annotations of different granularity, described below.
\textbf{Buckeye} \citep{pitt2005buckeye} contains interviews with hand-corrected word alignments. \textbf{Torgo} \citep{rudzicz2012torgo} covers atypical speech, with prompts read by speakers with dysarthria and hand phone alignments.
\textbf{DIHARD~III} \citep{ryant2021dihard3} and \textbf{VoxConverse} \citep{chung2020voxconverse} are diarization benchmarks, both scored with their own diarization labels: the former spans eleven domains, from audiobooks and telephone calls to clinical interviews and restaurant conversations, and the latter contains multi-speaker web video.
Robustness to noise is further evaluated on AMI with MUSAN noise \citep{snyder2015musan} added at 20 to 0\,dB SNR \citep{tan2020rvad,dinkel2021} and on \textbf{LibriVAD} \citep{stylianou2025librivad}, which mixes the read speech of LibriSpeech \citep{panayotov2015librispeech} with nine noise types at 20 to $-5$\,dB SNR.
Appendix~\ref{app:datasets} details their splits and references.

\begin{figure}[t]
\centering
\includegraphics[width=0.95\columnwidth]{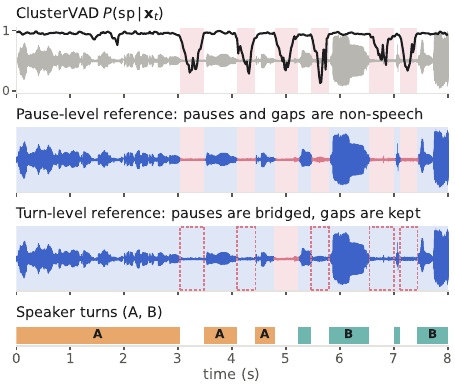}
\caption{Pause-level and turn-level references for an AMI excerpt with two speakers. The speech probability of \method{} (top) dips at every \nonspeechword{non-speech} interval of the pause-level reference; the turn-level reference labels the pauses within a turn as \speechword{speech} but keeps the gap between speakers.}
\label{fig:references}
\end{figure}

\begin{table*}[t]
\centering
\scriptsize
\setlength{\tabcolsep}{2.5pt}
\begin{tabular}{@{}lrrrrrrrrrrrr}
\toprule
 & \multicolumn{2}{c}{AMI} & \multicolumn{2}{c}{AMI far-field} & \multicolumn{2}{c}{AliMeeting} &  &  &  &  &  &  \\
\cmidrule(lr){2-3}\cmidrule(lr){4-5}\cmidrule(lr){6-7}
 & \multicolumn{1}{c}{pause-level} & \multicolumn{1}{c}{turn-level} & \multicolumn{1}{c}{pause-level} & \multicolumn{1}{c}{turn-level} & \multicolumn{1}{c}{pause-level} & \multicolumn{1}{c}{turn-level} & \multicolumn{1}{c}{Buckeye} & \multicolumn{1}{c}{Torgo} & \multicolumn{1}{c}{\smash{\makecell[b]{DIHARD\\III}}} & \multicolumn{1}{c}{\smash{\makecell[b]{Vox-\\Converse}}} & \multicolumn{1}{c}{\smash{\makecell[b]{AMI +\\MUSAN}}} & \multicolumn{1}{c}{LibriVAD} \\
\midrule
\textit{All speech (trivial)} & 51.8 & 24.3 & 51.8 & 24.3 & 20.5 & 8.7 & 37.1 & 140.6 & 26.4 & 11.7 & 51.8 & 18.9 \\
\midrule
\multicolumn{13}{@{}l}{\textit{Supervised}} \\
\quad pyannote 3.0 & 17.8{\tiny$\pm$2.6} & \textbf{6.2}{\tiny$\pm$1.2} & 20.6{\tiny$\pm$2.9} & \textbf{7.4}{\tiny$\pm$1.5} & 11.6{\tiny$\pm$2.3} & \textbf{3.8}{\tiny$\pm$0.8} & 15.4{\tiny$\pm$1.5} & 24.4{\tiny$\pm$2.2} & \textbf{7.9}{\tiny$\pm$0.6} & \textbf{5.0}{\tiny$\pm$0.4} & 19.5{\tiny$\pm$1.2} & 10.4{\tiny$\pm$0.2} \\
\quad Silero & 10.8{\tiny$\pm$1.1} & 7.9{\tiny$\pm$1.9} & 15.3{\tiny$\pm$2.5} & 11.0{\tiny$\pm$2.8} & 17.0{\tiny$\pm$5.8} & 10.0{\tiny$\pm$4.4} & 11.2{\tiny$\pm$1.5} & 17.3{\tiny$\pm$1.8} & 11.7{\tiny$\pm$1.3} & 5.7{\tiny$\pm$0.5} & 16.6{\tiny$\pm$1.7} & 10.9{\tiny$\pm$0.2} \\
\quad MarbleNet & 14.0{\tiny$\pm$1.1} & 8.9{\tiny$\pm$2.0} & 21.7{\tiny$\pm$2.9} & 12.7{\tiny$\pm$2.7} & 20.8{\tiny$\pm$5.0} & 9.3{\tiny$\pm$2.9} & 13.6{\tiny$\pm$1.5} & 18.1{\tiny$\pm$1.7} & 18.7{\tiny$\pm$1.7} & 5.6{\tiny$\pm$0.5} & 20.6{\tiny$\pm$1.8} & 12.5{\tiny$\pm$0.2} \\
\quad INA SSL-VAD & 10.6{\tiny$\pm$1.0} & 8.5{\tiny$\pm$1.9} & 16.7{\tiny$\pm$1.8} & 10.9{\tiny$\pm$2.1} & 13.6{\tiny$\pm$1.9} & 5.4{\tiny$\pm$1.1} & 9.2{\tiny$\pm$1.4} & 26.9{\tiny$\pm$2.5} & 13.0{\tiny$\pm$1.3} & 5.6{\tiny$\pm$0.5} & 18.3{\tiny$\pm$1.8} & 13.6{\tiny$\pm$0.3} \\
\midrule
\multicolumn{13}{@{}l}{\textit{Unsupervised}} \\
\quad rVAD & 16.8{\tiny$\pm$1.5} & 11.0{\tiny$\pm$1.7} & 19.7{\tiny$\pm$2.6} & 14.4{\tiny$\pm$2.9} & 14.5{\tiny$\pm$2.5} & 9.0{\tiny$\pm$2.7} & 17.4{\tiny$\pm$2.2} & 23.8{\tiny$\pm$1.9} & 13.9{\tiny$\pm$0.9} & 8.1{\tiny$\pm$0.8} & 24.5{\tiny$\pm$1.6} & 12.0{\tiny$\pm$0.2} \\
\quad de Jong \& Wempe & 22.0{\tiny$\pm$4.4} & 16.3{\tiny$\pm$3.1} & 40.4{\tiny$\pm$12.6} & 23.3{\tiny$\pm$6.0} & 15.6{\tiny$\pm$3.7} & 6.8{\tiny$\pm$1.9} & 11.7{\tiny$\pm$2.3} & 35.4{\tiny$\pm$4.5} & 16.3{\tiny$\pm$1.7} & 9.9{\tiny$\pm$1.0} & 29.9{\tiny$\pm$2.4} & 16.0{\tiny$\pm$0.2} \\
\quad WebRTC & 16.9{\tiny$\pm$2.4} & 11.9{\tiny$\pm$2.7} & 19.6{\tiny$\pm$2.3} & 14.5{\tiny$\pm$3.0} & 14.1{\tiny$\pm$2.9} & 6.5{\tiny$\pm$1.7} & 14.7{\tiny$\pm$2.2} & 24.2{\tiny$\pm$2.7} & 17.4{\tiny$\pm$1.5} & 9.3{\tiny$\pm$0.9} & 31.0{\tiny$\pm$2.5} & 13.7{\tiny$\pm$0.1} \\
\quad Energy & 15.9{\tiny$\pm$1.9} & 12.1{\tiny$\pm$2.6} & 24.8{\tiny$\pm$3.8} & 16.3{\tiny$\pm$2.7} & 11.5{\tiny$\pm$2.8} & 6.1{\tiny$\pm$2.3} & 12.0{\tiny$\pm$2.3} & 30.3{\tiny$\pm$3.4} & 13.6{\tiny$\pm$1.3} & 9.6{\tiny$\pm$0.9} & 27.9{\tiny$\pm$2.7} & 14.2{\tiny$\pm$0.1} \\
\midrule
\rowcolor{clustervadshade}
\multicolumn{13}{@{}l}{\textit{\method{}, tuned threshold}} \\
\rowcolor{clustervadshade}
\quad OLIVE-J & \textbf{8.1}{\tiny$\pm$0.8} & 7.5{\tiny$\pm$1.8} & \underline{11.0}{\tiny$\pm$1.7} & 9.6{\tiny$\pm$2.3} & 10.8{\tiny$\pm$2.4} & 5.9{\tiny$\pm$1.9} & \textbf{7.4}{\tiny$\pm$1.3} & \underline{14.2}{\tiny$\pm$1.4} & 8.6{\tiny$\pm$0.8} & 5.3{\tiny$\pm$0.5} & \textbf{15.1}{\tiny$\pm$1.7} & 8.8{\tiny$\pm$0.3} \\
\rowcolor{clustervadshade}
\quad data2vec~2.0 & 8.8{\tiny$\pm$1.0} & 9.0{\tiny$\pm$2.2} & 12.4{\tiny$\pm$1.6} & 10.7{\tiny$\pm$2.4} & 9.9{\tiny$\pm$1.7} & 5.5{\tiny$\pm$1.4} & 8.0{\tiny$\pm$1.4} & 17.4{\tiny$\pm$1.6} & 9.7{\tiny$\pm$1.0} & 6.2{\tiny$\pm$0.6} & 18.1{\tiny$\pm$1.8} & 8.3{\tiny$\pm$0.1} \\
\rowcolor{clustervadshade}
\quad HuBERT-L & 8.8{\tiny$\pm$0.9} & 9.3{\tiny$\pm$2.0} & 11.5{\tiny$\pm$1.4} & 11.9{\tiny$\pm$2.5} & \textbf{9.6}{\tiny$\pm$1.6} & 4.9{\tiny$\pm$1.0} & 7.6{\tiny$\pm$1.2} & 15.1{\tiny$\pm$1.9} & \underline{8.2}{\tiny$\pm$0.7} & 5.5{\tiny$\pm$0.5} & 15.4{\tiny$\pm$1.5} & \textbf{6.5}{\tiny$\pm$0.1} \\
\midrule
\rowcolor{clustervadbwshade}
\multicolumn{13}{@{}l}{\textit{\method{}, threshold-free decision}} \\
\rowcolor{clustervadbwshade}
\quad OLIVE-J & \underline{8.3}{\tiny$\pm$0.9} & \underline{7.2}{\tiny$\pm$1.8} & 11.2{\tiny$\pm$1.7} & 9.8{\tiny$\pm$2.2} & 11.6{\tiny$\pm$3.2} & 6.7{\tiny$\pm$2.6} & 8.3{\tiny$\pm$1.4} & \textbf{13.8}{\tiny$\pm$1.4} & 9.0{\tiny$\pm$1.0} & \underline{5.1}{\tiny$\pm$0.5} & \underline{15.3}{\tiny$\pm$1.7} & 7.8{\tiny$\pm$0.2} \\
\rowcolor{clustervadbwshade}
\quad data2vec~2.0 & 8.7{\tiny$\pm$1.0} & 8.5{\tiny$\pm$2.2} & 12.9{\tiny$\pm$1.6} & 11.4{\tiny$\pm$2.5} & 11.7{\tiny$\pm$2.8} & 6.6{\tiny$\pm$2.2} & \underline{7.6}{\tiny$\pm$1.6} & 22.3{\tiny$\pm$1.8} & 10.2{\tiny$\pm$1.0} & 6.2{\tiny$\pm$0.6} & 17.7{\tiny$\pm$1.8} & 10.5{\tiny$\pm$0.2} \\
\rowcolor{clustervadbwshade}
\quad HuBERT-L & 9.5{\tiny$\pm$1.2} & 8.4{\tiny$\pm$2.3} & \textbf{10.9}{\tiny$\pm$1.3} & \underline{9.5}{\tiny$\pm$2.1} & \underline{9.6}{\tiny$\pm$1.5} & \underline{4.7}{\tiny$\pm$1.0} & 8.7{\tiny$\pm$1.5} & 15.7{\tiny$\pm$2.5} & 10.5{\tiny$\pm$2.0} & 5.3{\tiny$\pm$0.5} & 15.8{\tiny$\pm$1.7} & \underline{7.3}{\tiny$\pm$0.1} \\
\bottomrule
\end{tabular}
\caption{DetER (\%, lower is better) on the benchmark test sets. Evaluation scores are reported with 95\% confidence intervals. Bold: best; underlined: second-best.}
\label{tab:main}
\end{table*}

\begin{figure}[t]
\centering
\includegraphics[width=0.93\columnwidth]{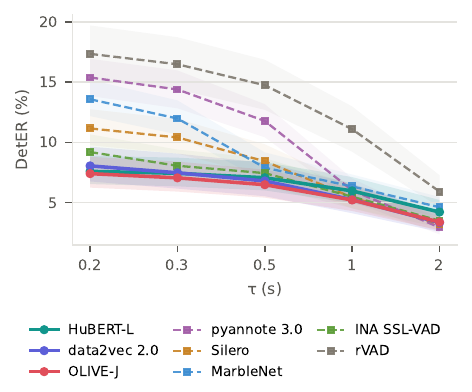}
\caption{DetER (\%) on Buckeye when pauses shorter than $\tau$ are labelled as speech in the reference ($\tau = 0.2$\,s: pause-level reference), with 95\% confidence intervals. Systems are tuned per reference.}
\label{fig:references-curve}
\end{figure}

\begin{figure*}[t]
\centering
\includegraphics[width=0.89\textwidth]{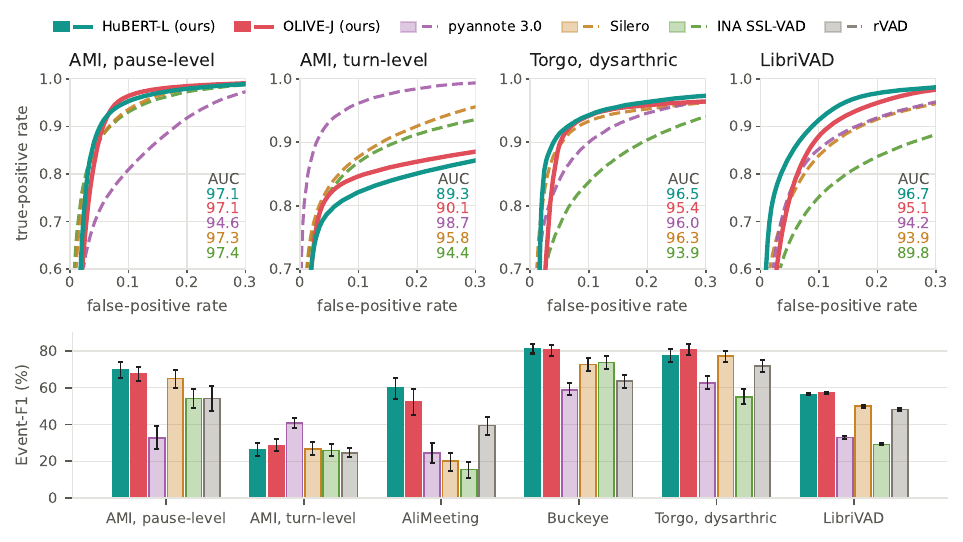}
\caption{Top: ROC curves and AUC where the systems differ. Bottom: Event-F1 with 95\% confidence intervals.}
\label{fig:roc-eventf1}
\end{figure*}

\paragraph{Reference annotations.}
\label{sec:references}
What counts as non-speech in a dataset depends on its annotation convention, which follows the intended use.
Following \citet{heldner2010}, we call a non-speech interval within one speaker's turn a \emph{pause} and one between two speakers a \emph{gap}.
References built for speech recognition typically keep semantically coherent stretches together and bridge the pauses inside a speaker's turn, whereas diarization references are finer: for instance, DIHARD splits a speaker's segment at pauses of 200\,ms or more \citep{ryant2021dihard3}.
A reference that bridges pauses penalises a detector for finding them: on AMI, for example, the manually timed words cover 90\% of the 0.2--2\,s pauses found by forced alignment almost entirely (Appendix~\ref{app:references}).
Supervised detectors learn the convention of their training labels and can carry it to other domains, whereas an unsupervised detector inherits none: it resolves pauses at the resolution of its features, and its segments must be merged across pauses when the target annotation convention requires it.
We therefore distinguish two kinds of references (Figure~\ref{fig:references}).
\emph{Pause-level} references keep every non-speech interval of 0.2\,s or more (both pauses and gaps). They typically come from forced alignment of manual transcripts. \citet{kraljevski2015} show that forced-alignment labels are as accurate as those of expert annotators, and such labels are widely used in VAD research \citep{tan2020rvad,bovbjerg2024} and in diarization benchmarks that need precise boundaries \citep{watanabe2020chime6,horiguchi2025}. For AMI and AliMeeting, we use the pause-level references of \citet{horiguchi2025}.
\emph{Turn-level} references follow speaker turns: they come from the manual annotations, bridge most pauses within a turn, and keep the gaps.

\paragraph{Metrics.}
\label{sec:metrics}
We evaluate detection at the frame level and at the segment level.
At the frame level, we report the detection error rate (DetER), which is the total duration of false positives (non-speech detected as speech) and false negatives (speech detected as non-speech) relative to the duration of reference speech \citep{bredin2017metrics}.
We complement it with the receiver operating characteristic (ROC) curve, which plots the true-positive rate against the false-positive rate as the threshold applied to a detector's frame-level speech scores varies. The area under this curve (AUC) summarises how well these scores separate speech from non-speech frames, independently of any decision.
At the segment level, we use Event-F1 \citep{mesaros2016}, the F1 score of the predicted speech segments, where a predicted segment matches a reference segment if their onsets differ by at most 200\,ms and their offsets by at most 200\,ms or 20\% of the reference duration \citep{dinkel2021}.

\paragraph{Tuning protocol.}
\label{sec:protocol}
Since annotation conventions differ, diarization toolkits re-tune detection thresholds and minimum durations for each dataset \citep{bredin2021segmentation,bredin2023pyannote}.
We likewise tune the decision of every system on the tuning data of each benchmark (Table~\ref{tab:datasets}), separately for each reference, to minimise DetER.
For the tuned threshold, the threshold is searched from 0.02 to 0.98 in steps of 0.02 and the minimum durations of speech and non-speech segments among 0, 0.1 and 0.2\,s; the three values are tuned jointly.
For baselines with discrete settings, such as the rVAD threshold or the WebRTC mode, the setting is part of the same search.
Finally, to match references that bridge pauses, duration merging relabels as speech every non-speech segment shorter than a minimum duration $\tau'$, chosen among 11 values from 0 to 5\,s for each target reference.
It is applied to every system, after either decision. For the threshold-free decision of \method{}, $\tau'$ is the only tuned value.

\section{Results and analysis}
\label{sec:results}

\begin{figure*}[t]
\centering
\includegraphics[width=0.89\textwidth]{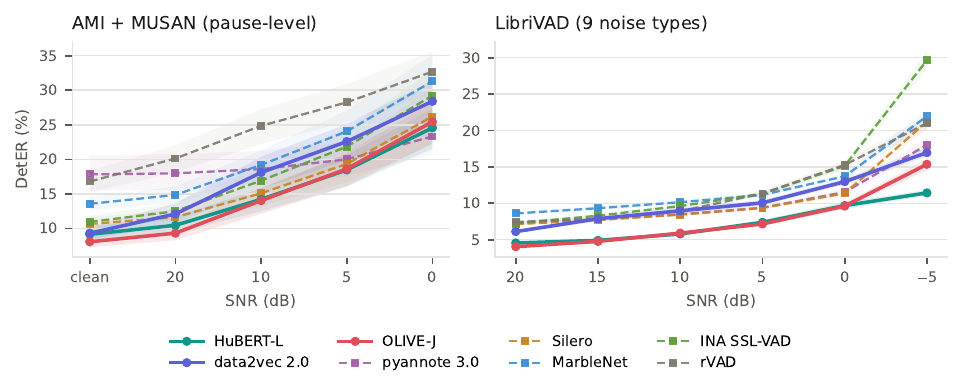}
\caption{DetER (\%) as a function of SNR, with 95\% confidence intervals. Each system uses a single decision, tuned over all SNRs; \method{} is plotted with the tuned threshold on AMI + MUSAN and the threshold-free decision on LibriVAD.}
\label{fig:noise}
\end{figure*}

\subsection{Detection error and reference granularity}
\label{sec:results-detection}
Table~\ref{tab:main} reports the detection error on the twelve benchmarks.
\method{} models obtain the lowest error on seven of them, which span close-talk and far-field meetings, Mandarin, interviews, dysarthric speech and noisy speech.
\method{}'s speech probability granularity resolves individual pauses, which suits pause-level references.
pyannote is strongest on the turn-level references and on the two diarization benchmarks; it is trained on data from these corpora, and they are therefore in domain for it which plays to the strength of the supervised approach.
\method{} still performs well there: it obtains the second-lowest error on all of them and stays within the confidence intervals of pyannote.
\method{} models also outperform the unsupervised baselines.
Moreover, \method{} with HuBERT-L and the threshold-free decision outperforms INA SSL-VAD, its supervised counterpart, on every benchmark.
With Base encoders, as in INA SSL-VAD, it does so on eleven benchmarks with OLIVE-J and on nine with data2vec~2.0.

To isolate the effect of reference granularity, Figure~\ref{fig:references-curve} shows the detection error on Buckeye as pauses shorter than $\tau$ are closed in the reference.
When the reference keeps short pauses, \method{} is ahead of every baseline.
As $\tau$ grows, the error of the supervised detectors falls and the neural systems converge, while that of \method{} varies the least: it resolves pauses, and duration merging adapts it to coarser references.

\subsection{Threshold-free and tuned decisions}
\label{sec:results-decisions}
Table~\ref{tab:main} also compares the two decisions of \method{}.
With OLIVE-J, the threshold-free decision yields a lower error than the tuned threshold or lies within its reported confidence interval on every benchmark.
For the other encoders, the decisions differ mainly on Torgo and LibriVAD, whose short utterances may make fitting the threshold-free decision less reliable.

\subsection{ROC analysis and segment detection}
\label{sec:results-scores}
Figure~\ref{fig:roc-eventf1} first shows the ROC curves and AUC of the frame-level speech scores, which do not depend on the decision.
On pause-level AMI and on dysarthric speech, the AUC of \method{} is close to that of the best supervised detectors.
The curves show that the supervised detectors achieve higher true-positive rates at low false-positive rates, whereas \method{} achieves higher true-positive rates at moderate false-positive rates.
These differences in operating points help explain the difference in DetER.
On turn-level AMI, the scores of \method{} separate speech from non-speech less well than those of pyannote, because they dip at the pauses that this reference labels as speech; duration merging, which the ROC curve does not capture, makes up for this difference at the decision stage which enables \method{} to perform well in terms of DetER.
On LibriVAD, \method{} with HuBERT-L has the highest AUC.

The Event-F1 of the predicted segments is also reported in Figure~\ref{fig:roc-eventf1}. \method{} obtains the highest Event-F1 on all the benchmarks shown except turn-level AMI, indicating that its segment boundaries are the closest to those of the references.

\subsection{Robustness to noise}
\label{sec:results-noise}
Figure~\ref{fig:noise} examines how detection error changes as SNR decreases.
\method{} has the lowest error at every SNR on LibriVAD and at every SNR except 0\,dB on AMI + MUSAN, where pyannote is slightly lower, within the confidence intervals. We note that LibriVAD is in domain for the SSL encoders, which are pre-trained on audiobook speech.
\section{Discussion and conclusions}
\label{sec:conclusion}
In this work, we demonstrated that speech against non-speech information is inherently captured in frozen SSL representations, and that we can explicitly leverage this information through clustering, allowing voice activity detection that inherits the robustness and generalisation ability of the representations without training a detector.
Indeed, although the PCA whitening, the codebook and the cluster labels were fitted once, on only 1.8\,h of clean, read English speech, and the layer of each encoder was chosen on a single development set, \method{} generalised across domains.
This one configuration was applied unchanged to meetings, far-field and noisy recordings, Mandarin and dysarthric speech, where it obtained the lowest or second-lowest error against detectors trained on labelled speech. This highlights the robustness of the representations.
The choice of layer is not critical either: all but a few layers of each encoder performed close to its best layer (Appendix~\ref{app:layers}), which suggests that the distinction between speech and non-speech is carried throughout the encoder.
Finally, labelling the clusters with sonority, a form of self-supervision by an acoustic measure, was sufficient: while sonority is a weak detector on its own and degrades quickly in noisy conditions (Appendix~\ref{app:naming}), the clusters it labels yield a speech probability that is itself robust.

On turn-level references, pyannote still performed best.
Besides being in domain for these benchmarks, it is a speaker diarization model, which can leverage speaker information to tell a pause within a turn from a change of speaker. Meanwhile, \method{} bridges pauses by their duration only.
Future work will investigate integrating speaker information to enable speaker-aware pause bridging.
We further showed that the threshold-free decision matched the tuned threshold across benchmarks, removing the need to tune a threshold.
However, it is less reliable on short utterances, possibly because less audio is available to fit the model. Future work will explore ways to make it more robust to such inputs.

Overall, \method{} reconciles the cross-domain generalisation of knowledge-driven unsupervised VAD with the ability of supervised VAD to benefit from data, leveraging frozen SSL representations learned from large corpora of untranscribed speech.
\method{} combines these representations, a small codebook fitted on unlabelled auxiliary data, and sonority as an acoustic measure to achieve unsupervised VAD that is competitive with supervised detectors across domains.

\section*{Limitations}
\paragraph{Computational efficiency.}
\method{} inherits the computational cost of its SSL encoder, which exceeds that of compact detectors.
This cost is nevertheless limited, since the encoder is only needed up to the selected layer (Appendix~\ref{app:cost}). For instance, the data2vec~2.0 variant only uses the convolutional feature encoder.
Nevertheless, building the codebook on lighter representations, such as distilled encoders, is left to future work.

\paragraph{Input duration.}
Since the threshold-free decision is fitted on each recording, it is less reliable on short inputs, where little audio is available for the fit, and it falls behind the tuned threshold for some encoders.

\paragraph{Speaker-aware pause merging.}
References built for speech recognition typically bridge the pauses between speech segments of the same speaker.
Diarization systems can bridge such pauses because they integrate speaker information into the detection, whereas \method{} merges segments by their duration only.
Integrating speaker information into the pipeline would allow \method{} to adapt its detection to such references.

\paragraph{Languages.}
The codebook is built from English speech, and the benchmarks cover English and Mandarin only.
\method{} transfers to Mandarin without any change, but its performance on other languages remains to be evaluated.

\section*{Ethical considerations}
This work uses existing research corpora under their respective licences and collects no new data.
Voice activity detectors are an essential component of speech processing pipelines, and they help enable and advance systems such as speech language models, speech recognisers and speech synthesisers.
Speech technology is valuable in many domains, but it can also be misused, for example for monitoring or voice cloning.
Raising awareness of these risks and advancing watermarking and deepfake detection are essential to limit malicious use.

\section*{Acknowledgements}
This work was partially supported by the Swiss National Science Foundation through the project ``Pathological Speech Synthesis (PaSS)'', grant agreement no. 219726.

\bibliography{custom}

\appendix
\section{Encoders and layers}
\label{app:representation}

\subsection{Encoder and layer selection}
\label{app:layers}
\begin{figure}[!htb]
\centering
\includegraphics[width=\columnwidth]{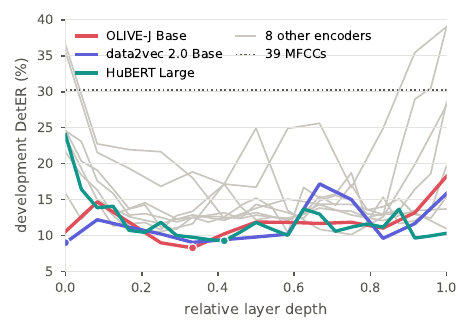}
\caption{DetER (\%) on the AMI development set (pause-level reference) across the layers of 11 encoders.}
\label{fig:app-layers}
\end{figure}

We run \method{} on every layer of 11 frozen encoders, Base and Large: wav2vec~2.0 \citep{baevski2020wav2vec}, XLSR-53 \citep{conneau2021xlsr}, HuBERT \citep{hsu2021hubert}, WavLM \citep{chen2022wavlm}, data2vec \citep{baevski2022data2vec}, data2vec~2.0 \citep{baevski2023data2vec2} and OLIVE-J \citep{elhajal2026olive}, and select the encoder and layer on AMI development data with the pause-level reference.
OLIVE-J at layer 4, data2vec~2.0 at layer 0 and HuBERT Large at layer 10 obtain the three lowest errors (Table~\ref{tab:app-layers}).
For these encoders, most layers perform close to the best one, and every layer outperforms the same pipeline on MFCCs (Figure~\ref{fig:app-layers}).

\begin{table}[!htb]
\centering
\footnotesize
\setlength{\tabcolsep}{4pt}
\begin{tabular}{@{}llrrr@{}}
\toprule
Encoder & Size & Best layer & DetER & Median \\
\midrule
\rowcolor{clustervadshade}OLIVE-J & Base & 4 of 12 & 8.3 & 11.8 \\
\rowcolor{clustervadshade}data2vec~2.0 & Base & 0 of 12 & 9.0 & 10.2 \\
\rowcolor{clustervadshade}HuBERT & Large & 10 of 24 & 9.3 & 10.9 \\
data2vec & Base & 9 of 12 & 10.1 & 12.6 \\
data2vec & Large & 14 of 24 & 10.5 & 13.9 \\
WavLM & Large & 7 of 24 & 10.9 & 13.0 \\
WavLM Base+ & Base & 12 of 12 & 11.0 & 13.2 \\
wav2vec~2.0 & Large & 7 of 24 & 11.9 & 13.8 \\
HuBERT & Base & 10 of 12 & 12.1 & 13.7 \\
XLSR-53 & Large & 12 of 24 & 12.2 & 13.5 \\
wav2vec~2.0 & Base & 9 of 12 & 16.7 & 21.5 \\
\midrule
39 MFCCs & -- & -- & 30.2 & -- \\
\bottomrule
\end{tabular}
\caption{DetER (\%) on the AMI development set (pause-level reference) at the best layer of each encoder, and its median over layers.}
\label{tab:app-layers}
\end{table}

\subsection{Hand-crafted features}
\label{app:mfcc}
To assess the contribution of the SSL representations, we run the \method{} pipeline unchanged on 39 MFCCs.
The results, shown in Table~\ref{tab:app-mfcc}, highlight the importance of the representations for the method.

\begin{table}[!htb]
\centering
\scriptsize
\setlength{\tabcolsep}{2.4pt}
\begin{tabular}{@{}lrrrrrrrr@{}}
\toprule
 & \multicolumn{2}{c}{OLIVE-J} & \multicolumn{2}{c}{data2vec~2.0} & \multicolumn{2}{c}{HuBERT-L} & \multicolumn{2}{c}{MFCC} \\
\cmidrule(lr){2-3}\cmidrule(lr){4-5}\cmidrule(lr){6-7}\cmidrule(lr){8-9}
 & thr. & free & thr. & free & thr. & free & thr. & free \\
\midrule
AMI, pause-level & 7.6 & 7.7 & 7.9 & 7.8 & 8.2 & 8.6 & 29.6 & 13.9 \\
AMI, turn-level & 7.6 & 7.4 & 8.4 & 8.1 & 9.2 & 8.3 & 17.9 & 10.2 \\
AMI far-field & 9.7 & 9.8 & 11.1 & 11.4 & 10.4 & 9.9 & 32.1 & 18.6 \\
AliMeeting & 10.9 & 11.8 & 10.6 & 12.1 & 10.0 & 9.9 & 11.5 & 19.7 \\
Buckeye & 7.4 & 8.0 & 6.9 & 7.1 & 7.6 & 8.3 & 9.4 & 7.9 \\
VoxConverse & 3.2 & 3.2 & 3.7 & 3.6 & 3.3 & 3.3 & 5.9 & 6.1 \\
\bottomrule
\end{tabular}
\caption{DetER (\%) on the development sets for \method{} with SSL features and with 39 MFCCs, with the tuned threshold (thr.) and the threshold-free decision (free).}
\label{tab:app-mfcc}
\end{table}

\subsection{Inference window and layer combination}
Table~\ref{tab:app-options} varies the inference window and combines three adjacent layers by averaging or concatenation.
Longer windows perform like the 5\,s windows of the paper and shorter ones slightly worse; combining layers does not help.

\begin{table}[!htb]
\centering
\scriptsize
\setlength{\tabcolsep}{3pt}
\begin{tabular}{@{}lrrr@{}}
\toprule
 & OLIVE-J & data2vec~2.0 & HuBERT Large \\
\midrule
\multicolumn{4}{@{}l}{\emph{Inference window}} \\
\quad 2\,s & 9.1 & 9.2 & 10.4 \\
\quad 5\,s (paper) & 8.3 & 9.0 & 9.3 \\
\quad 10\,s & 8.2 & 9.0 & 8.9 \\
\quad 20\,s & 8.1 & 9.0 & 8.7 \\
\multicolumn{4}{@{}l}{\emph{Layers}} \\
\quad 1 layer (paper) & 8.3 & 9.0 & 9.3 \\
\quad 3 layers, mean & 8.3 & 9.7 & 9.9 \\
\quad 3 layers, concatenated & 9.8 & 9.5 & 9.5 \\
\bottomrule
\end{tabular}
\caption{DetER (\%) on the AMI development set (pause-level reference) for different inference windows and for combinations of three adjacent layers.}
\label{tab:app-options}
\end{table}

\section{PCA whitening}
\label{app:anisotropy}
\begin{figure}[!htb]
\centering
\includegraphics[width=\columnwidth]{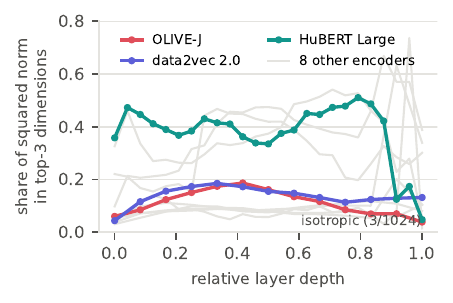}
\caption{Share of the squared feature norm held by the three largest coordinates, per layer of 11 encoders.}
\label{fig:anisotropy}
\end{figure}

SSL features are anisotropic: a few coordinates hold a large share of the squared norm of a frame, especially in the Large encoders (Figure~\ref{fig:anisotropy}), which can bias cosine similarity towards these coordinates.
PCA whitening, the fitting step \fitstep{A}, removes this effect and lowers the development error (Table~\ref{tab:normalisation}).
Table~\ref{tab:pca-variance} gives the share of the variance that the retained components keep.

\begin{table}[!htb]
\centering
\small
\begin{tabular}{@{}lc@{}}
\toprule
Normalisation & dev \\
\midrule
none & 11.8 \\
PCA whitening, $d=64$ & 8.6 \\
PCA whitening, $d=128$ & \textbf{8.3} \\
\bottomrule
\end{tabular}
\caption{DetER (\%) on the AMI development set (pause-level reference) for OLIVE-J, without and with PCA whitening.}
\label{tab:normalisation}
\end{table}

\begin{table}[!htb]
\centering
\small
\begin{tabular}{@{}lccc@{}}
\toprule
Encoder (layer) & $D$ & $d=64$ & $d=128$ \\
\midrule
OLIVE-J (4) & 768 & 74.9 & 85.9 \\
data2vec~2.0 (0) & 768 & 70.9 & 85.5 \\
HuBERT Large (10) & 1024 & 66.5 & 76.8 \\
\midrule
Base encoders, all layers & 768 & 36--75 & 54--86 \\
HuBERT Large, all layers & 1024 & 46--89 & 62--96 \\
\bottomrule
\end{tabular}
\caption{Variance (\%) of the codebook data retained by the $d$ leading principal components ($D$: feature dimension).}
\label{tab:pca-variance}
\end{table}

\section{Codebook ablations}
\label{app:codebook}
Table~\ref{tab:codebook} ablates the codebook of OLIVE-J on the AMI development set, one setting at a time.
The detection error is largely insensitive to the source of the codebook data, to its amount from 100 to 10{,}000 utterances, to the number of clusters between 50 and 200, and to the assignment (temperature between 0.05 and 0.2, cosine or Euclidean distance).
$K$-means and spherical $k$-means are equivalent.
The error is also stable across data subsets and random seeds.

\begin{table}[!htb]
\centering
\footnotesize
\setlength{\tabcolsep}{3pt}
\begin{tabular}{@{}llcc@{}}
\toprule
Factor & Setting & pause-level & turn-level \\
\midrule
Codebook data & LJSpeech & \textbf{8.3} & \textbf{15.2} \\
(1k utt.) & LibriSpeech & 8.4 & 16.0 \\
 & AMI train & 10.8 & 15.6 \\
 & mixture of the three & 8.9 & 15.6 \\
\midrule
LJSpeech size & 100 utt. & 9.3 & 16.8 \\
 & 1k utt. & \textbf{8.3} & \textbf{15.2} \\
 & 10k utt. & 9.1 & 16.8 \\
\midrule
Clusters $K$ & 2 & 46.5 & 27.3 \\
 & 10 & 11.1 & 16.4 \\
 & 50 & 8.9 & 15.8 \\
 & 100 & \textbf{8.3} & \textbf{15.2} \\
 & 200 & 9.0 & 17.2 \\
 & 500 & 9.6 & 17.9 \\
\midrule
Clustering & $k$-means & \textbf{8.3} & \textbf{15.2} \\
 & spherical $k$-means & 8.7 & 16.5 \\
\midrule
Assignment & cosine, $T=0.05$ & 8.5 & \textbf{15.0} \\
 & cosine, $T=0.1$ & \textbf{8.3} & 15.2 \\
 & cosine, $T=0.2$ & 8.6 & 15.6 \\
 & cosine, $T=0.5$ & 9.9 & 16.8 \\
 & Euclidean, $T=0.5$ & 8.5 & 15.3 \\
 & Euclidean, $T=1$ & 9.6 & 16.6 \\
\midrule
Seeds & 3 subsets $\times$ 3 seeds & 8.5 & 16.2 \\
 & (standard deviation) & $\pm$0.4 & $\pm$0.6 \\
\bottomrule
\end{tabular}
\caption{Codebook ablations for OLIVE-J: DetER (\%) on the AMI development set. One setting varies per block; bold: the configuration of the paper.}
\label{tab:codebook}
\end{table}

\section{Cluster labelling}
\label{app:naming}
Table~\ref{tab:app-naming} compares five signals for labelling the clusters: sonority, frame energy, the SSL feature norm, WebRTC VAD decisions, and the Urhythmic grouping \citep{vanniekerk2023urhythmic}.
Sonority obtains the lowest or tied-lowest development error for the three encoders, without relying on another detector.

\begin{table}[!htb]
\centering
\footnotesize
\setlength{\tabcolsep}{2pt}
\begin{tabular}{@{}lrrr@{}}
\toprule
Labelling signal & OLIVE-J & data2vec~2.0 & HuBERT Large \\
\midrule
Urhythmic groups & 20.2\,{\scriptsize(L5)} & 12.7\,{\scriptsize(L7)} & 12.3\,{\scriptsize(L6)} \\
WebRTC VAD rate & 9.8\,{\scriptsize(L3)} & 10.0\,{\scriptsize(L0)} & 9.3\,{\scriptsize(L8)} \\
frame energy & 11.8\,{\scriptsize(L9)} & 9.7\,{\scriptsize(L6)} & 14.1\,{\scriptsize(L14)} \\
\rowcolor{clustervadshade}sonority & 8.3\,{\scriptsize(L4)} & 9.0\,{\scriptsize(L0)} & 9.3\,{\scriptsize(L10)} \\
feature norm & 10.2\,{\scriptsize(L4)} & 10.6\,{\scriptsize(L3)} & 11.7\,{\scriptsize(L23)} \\
\bottomrule
\end{tabular}
\caption{DetER (\%) on the AMI development set (pause-level reference) for each labelling signal, at its best layer (in brackets).}
\label{tab:app-naming}
\end{table}

\paragraph{Sonority as a detector.}
Used directly as the speech score, sonority is clearly worse than \method{} on clean speech and degrades faster in noisy conditions (Table~\ref{tab:app-sonority}).

\begin{table*}[t]
\centering
\scriptsize
\setlength{\tabcolsep}{3pt}
\begin{tabular}{@{}lrrrrrrrr}
\toprule
 & \multicolumn{4}{c}{DetER (\%, lower is better)} & \multicolumn{4}{c}{AUC (\%, higher is better)} \\
\cmidrule(lr){2-5}\cmidrule(l){6-9}
 & \multicolumn{2}{c}{AMI + MUSAN} & \multicolumn{2}{c}{LibriVAD} & \multicolumn{2}{c}{AMI + MUSAN} & \multicolumn{2}{c}{LibriVAD} \\
 & clean & 0\,dB & 20\,dB & $-5$\,dB & clean & 0\,dB & 20\,dB & $-5$\,dB \\
\midrule
sonority & 20.8{\tiny$\pm$2.7} & 37.5{\tiny$\pm$5.7} & 7.4{\tiny$\pm$0.2} & 18.4{\tiny$\pm$0.5} & 91.8{\tiny$\pm$1.7} & 74.7{\tiny$\pm$2.8} & 96.7{\tiny$\pm$0.1} & 72.6{\tiny$\pm$0.7} \\
frame energy & 15.8{\tiny$\pm$1.7} & 39.7{\tiny$\pm$6.1} & 8.2{\tiny$\pm$0.2} & 19.3{\tiny$\pm$0.4} & 90.6{\tiny$\pm$2.0} & 69.2{\tiny$\pm$2.1} & 93.3{\tiny$\pm$0.1} & 63.2{\tiny$\pm$0.4} \\
\rowcolor{clustervadshade}\method{} OLIVE-J & 8.1{\tiny$\pm$0.8} & 25.4{\tiny$\pm$3.0} & 3.9{\tiny$\pm$0.1} & 22.1{\tiny$\pm$1.3} & 97.1{\tiny$\pm$0.5} & 90.6{\tiny$\pm$1.8} & 98.0{\tiny$\pm$0.1} & 86.6{\tiny$\pm$0.8} \\
\rowcolor{clustervadshade}\method{} HuBERT-L & 9.2{\tiny$\pm$1.0} & 24.6{\tiny$\pm$2.9} & 4.2{\tiny$\pm$0.1} & 10.3{\tiny$\pm$0.3} & 97.1{\tiny$\pm$0.4} & 91.1{\tiny$\pm$1.6} & 98.4{\tiny$\pm$0.1} & 89.6{\tiny$\pm$0.7} \\
\bottomrule
\end{tabular}
\caption{DetER and AUC (\%) of sonority and frame energy used directly as speech scores, and of \method{}, in the cleanest and noisiest condition of each noise benchmark.}
\label{tab:app-sonority}
\end{table*}

\section{Baselines}
\label{app:baselines}
Each baseline is run in its default configuration, and its decision is tuned on each benchmark (Section~\ref{sec:protocol}). For the unsupervised baselines, discrete settings are included in this search.

\paragraph{Supervised.}
For pyannote segmentation~3.0, the speech score is one minus the probability that no speaker is active; the model is trained on AISHELL, AliMeeting, AMI, AVA-AVD, DIHARD, Ego4D, MSDWild, REPERE and VoxConverse.
The training data of Silero VAD are not public.
MarbleNet (frame-level multilingual v2.0) is trained on about 3{,}900\,h that include AMI, ICSI, LibriSpeech, Fisher and Multilingual LibriSpeech.
For INA SSL-VAD, we take the frame probabilities before its Viterbi decoding; its encoder is pre-trained on French TV and radio data and its classifier trained on InaGVAD \citep{doukhan-etal-2024-inagvad}.

\paragraph{Unsupervised.}
For rVAD, we use the authors' MATLAB implementation and search over its threshold (0.2, 0.4 and 0.6).
For the detector of \citet{dejong2009}, we search over the minimum pause duration and silence level (the default 0.3\,s at $-25$\,dB, and 0.1\,s at $-20$, $-25$ and $-30$\,dB).
For WebRTC VAD, we search over its four modes (0 to 3).
For the frame-energy detector, we search over per-recording and global scaling of the frame log-energy.

\makeatletter
\setlength{\@fptop}{0pt}
\setlength{\@fpsep}{\floatsep}
\makeatother
\begin{table}[t]
\centering
\footnotesize
\setlength{\tabcolsep}{4pt}
\begin{tabular}{@{}lrrrr@{}}
\toprule
 & \multicolumn{2}{c}{AMI} & \multicolumn{2}{c}{AliMeeting} \\
\cmidrule(lr){2-3}\cmidrule(l){4-5}
 & pause & turn & pause & turn \\
\midrule
Speech share (\%) & 65.9 & 80.4 & 83.0 & 92.0 \\
Speech segments per minute & 22.2 & 5.6 & 20.2 & 7.1 \\
Median segment (s) & 1.24 & 3.64 & 1.77 & 4.13 \\
Pauses per minute & 14.0 & 2.0 & 11.8 & 2.7 \\
Gaps per minute & 8.1 & 3.6 & 8.4 & 4.3 \\
Median non-speech (s) & 0.48 & 1.10 & 0.37 & 0.49 \\
\bottomrule
\end{tabular}
\caption{Statistics of the pause-level and turn-level references of the AMI and AliMeeting test sets. A pause separates two speech segments of one speaker; a gap separates two speakers.}
\label{tab:app-reference-stats}
\end{table}

\begin{table}[t]
\centering
\footnotesize
\setlength{\tabcolsep}{3pt}
\begin{tabular}{@{}lrrrrr@{}}
\toprule
 & Params & \multicolumn{2}{c}{RTF} & \multicolumn{2}{c}{Memory (GB)} \\
\cmidrule(lr){3-4}\cmidrule(l){5-6}
Encoder (layers) & (M) & CPU & GPU & RAM & GPU \\
\midrule
OLIVE-J (0--4) & 36.5 & 0.27 & 0.0015 & 4.2 & 1.7 \\
data2vec~2.0 (0) & 8.1 & 0.19 & 0.0011 & 3.5 & 1.6 \\
HuBERT-L (0--10) & 139.1 & 0.56 & 0.0026 & 4.4 & 2.3 \\
\bottomrule
\end{tabular}
\caption{Encoder parameters up to the selected layer, real-time factor (RTF, processing time over audio duration; lower is faster) and peak memory of \method{} on 30 minutes of AMI audio, on one CPU thread or one RTX~3090 GPU.}
\label{tab:app-cost}
\end{table}

\section{Datasets}
\label{app:datasets}
This appendix details the benchmarks of Table~\ref{tab:datasets}.

\paragraph{Meetings.}
For AMI \citep{carletta2005ami}, we use the standard diarization splits and the word-based turn-level reference of \citet{landini2022vbx}, and the forced-alignment pause-level reference of \citet{horiguchi2025}; the far-field version uses the first microphone of the array.
For AliMeeting \citep{yu2022m2met}, we use the pause-level reference of \citet{horiguchi2025} and the original annotation as turn-level reference.

\paragraph{Noise.}
MUSAN noise \citep{snyder2015musan} is added to the AMI headset mix at 20, 10, 5 and 0\,dB SNR, and one decision is tuned on the development set pooled over clean audio and every SNR.
For LibriVAD \citep{stylianou2025librivad}, we use the NonConcat version, with its medium test set and its small validation set for tuning.

\paragraph{Buckeye.}
We score the participant's speech against the hand-corrected word alignments \citep{pitt2005buckeye} and leave the interviewer's turns unscored, with the speaker split of \citet{kamper2017segmental}.

\paragraph{Torgo.}
We evaluate on the three dysarthric speakers for which hand phone alignments are available \citep{rudzicz2012torgo}. We use a Leave-One-Speaker-Out approach for tuning, where each speaker is scored with the decision tuned on the other two.

\paragraph{Diarization benchmarks.}
For DIHARD~III \citep{ryant2021dihard3}, we use the evaluation set with its official references, split into two folds stratified by domain, each scored with the decision tuned on the other.
For VoxConverse \citep{chung2020voxconverse}, we tune on the development set and test on the test set.

\paragraph{Confidence intervals.}
All confidence intervals are 95\% percentile bootstrap intervals over the test files (1{,}000 resamples).

\section{Reference annotations}
\label{app:references}
Table~\ref{tab:app-reference-stats} compares the pause-level and turn-level references of AMI and AliMeeting.
The turn-level references bridge most pauses within a turn and keep about half of the gaps between speakers. They therefore label more of the audio as speech and have fewer and longer segments.
On AMI, the manually timed words of the turn-level reference cover 90\% of the 0.2--2\,s pauses of the pause-level reference almost entirely, so a detector that finds these pauses is penalised.

\section{Computational cost}
\label{app:cost}
Table~\ref{tab:app-cost} reports encoder size and inference cost when extraction stops at the selected layer.

\end{document}